\documentclass[journal,12pt]{IEEEtran}

\usepackage[pdftex]{graphicx}
\usepackage{amsmath}
\usepackage{eqparbox}
\usepackage{xcolor}
\usepackage{caption}
\usepackage{subcaption}
\usepackage{amssymb}
\usepackage{mathtools}
\usepackage{multirow}
\usepackage[utf8]{inputenc}
\usepackage{float}
\usepackage{booktabs}
\usepackage{fancyhdr}
\usepackage{algorithm}
\usepackage[noend]{algpseudocode}

\begin{document}


\title{\LARGE Underwater Acoustic Communication Receiver Using Deep Belief Network}
\author{\authorblockN{Abigail Lee-Leon\authorrefmark{1}\authorrefmark{2}, Chau Yuen\authorrefmark{1}, Dorien Herremans\authorrefmark{1}}
\authorblockA{\authorrefmark{1}Singapore University of Technology and Design (SUTD), 8 Somapah Road, Singapore 487372}
\authorblockA{\authorrefmark{2}Thales Solutions Asia Pte Ltd, 21 Changi North Rise, Singapore 498788}}
\maketitle

\begin{abstract}
Underwater environments create a challenging channel for communications. In this paper, we design a novel receiver system by exploring the machine learning technique--Deep Belief Network (DBN)-- to combat the signal distortion caused by the Doppler effect and multi-path propagation. We evaluate the performance of the proposed receiver system in both simulation experiments and sea trials. Our proposed receiver system comprises of DBN based de-noising and classification of the received signal. First, the received signal is segmented into frames before the each of these frames is individually pre-processed using a novel pixelization algorithm. Then, using the DBN based de-noising algorithm, features are extracted from these frames and used to reconstruct the received signal. Finally, DBN based classification of the reconstructed signal occurs. Our proposed DBN based receiver system does show better performance in channels influenced by the Doppler effect and multi-path propagation with a performance improvement of 13.2dB at $10^{-3}$ Bit Error Rate (BER).
\end{abstract}


\IEEEpeerreviewmaketitle

\section{Introduction}

Underwater Acoustic Communications (UWAC) is a knowledge rich field that has, in the recent years, gained a tremendous amount of interest for its many applications in the field of ocean exploration, defense, and marine commercial industries. A few notable applications are underwater exploration \cite{Sujit}, underwater mine detection \cite{Yeu}, and underwater communications between submarines or underwater nodes \cite{Al-Dharrab}. Due to a rapidly growing need for data-heavy underwater systems, the expectations and requirements of the underwater system design has risen up to a point where a growing number of researchers are starting to turn to unconventional methods like machine learning (ML) and deep learning (DL) to combat the challenging underwater environment. In this paper, we propose a novel receiver system that utilizes the capabilities of Deep Belief Networks (DBNs) to redesign the de-noising and demodulation blocks of the communication system. 

Generally, conventional signal processing algorithms in communications are based on strong mathematical foundations and are designed specifically for a variety of specific channels and system models \cite{Rappaport,Almers2007}. For instance, the Binary Phase-Shift Key (BPSK) modulation was designed for the detection of a constellation symbol in a channel of additive white Guassian noise (AWGN) \cite{L.Hong2000}. These signal processing algorithms are constructed on expert knowledge of the tractable channel models, which in turn are established on a simplification of Maxwell's equations \cite{F.K.Jondral2008}. 

UWAC signals, however, are not electromagnetic in nature \cite{M.Stojanovic2009,Yusof2012}. As such, the UWAC channel is widely characterized as one of the most complex channels to model and has yet to develop a palpable or definite model. Its high complexity is mostly derived from its fast varying characteristics, such as the Doppler effect and the propagation properties. 

Since communications are heavily reliant on the characteristics of sound, it shows a strong correlation with the properties of sound \cite{M.Stojanovic2009}. By understanding how sounds are influenced during sea trials, one can optimize the efficiency of the communication through adaptation. As sound propagates underwater at a very low speed of approximately 1500 m/s \cite{Yusof2012}, and propagation occurs over multiple paths, it is very common to observe a delay spreading over tens or even hundreds of milliseconds which results in frequency-selective signal distortion. This motion also results in an extreme Doppler effect. 

Multi-path propagation in the ocean is governed by three effects-- (1) sound reflecting off underwater surfaces like bubbles and the seabed, (2) sound refraction in the water due to density change, and (3) energy loss \cite{Urick1975}. These effects will cause an elongation of the path traveled, and thus a time delay. The first also creates reverberation, which causes a reflection phase change and a reflection amplitude change. The second is a consequence of the spatial variability of sound speed, which is dependent on temperature, salinity, and pressure. These factors vary with depth and location. The final effect is heavily dependent on the signal frequency, as well as the pH level of the water. This dependence is a consequence of absorption, where the signal energy is converted to heat. In addition to the absorption loss, the signal typically experiences a spreading loss, which increases with distance \cite{M.Stojanovic2009}. To correct for the intersymbol interference (ISI) caused by the propagation, the works done in \cite{Stojanovic1994,Stojanovic1996,Kilfoyle2000} used an adaptive equalizer to flexibly compensate for the changes in the channel. 

Another distinguishing property of UWAC is the channel's time variability -- (1) inherent changes in the propagation medium and (2) transmitter/receiver motion. Inherent changes include long term changes like seasonal temperatures and instantaneous changes caused by shipping routes and moving water surfaces. These factors result in both a scattering of the signal and a Doppler effect spreading due to the changing path length \cite{Urick1975}. A combination of these factors creates a complex challenge of modeling a sufficiently accurate channel model. To combat Doppler shifts, many Doppler scale estimation techniques have been proposed as seen in \cite{wan2012performance, YC}.

In recent years, ML and more specifically DL have gained recognition for their performance in fields known for their high modelling complexity~\cite{Zheng2010}, such as image recognition~\cite{Mashford}, natural language processing~\cite{Herremans2019}, and handwriting analysis \cite{M.M.A.Ghosh}. Currently researchers have begun to explore the applications of ML and DL in the area of communications. For example, recent research by Wang et al.~\cite{Wang} exploited deep learning to detect signal modulations in underwater channels. Other studies like \cite{V.Q.Dang,T.O'Shea} used deep neural network (NN)-based auto encoders to demodulate received signals. As such, we can expect that applying ML techniques to communication blocks in order to provide a promising solution to the complex channel problem will yield significant improvements in decoding the physical layer. Works in \cite{Youwen2019,Wang2019IEEE} investigate the use of DL based orthogonal frequency-division multiplexing (OFDM) receiver to recover signals corrupted by the UWAC channel. 

In this paper, we explore a receiver system that fundamentally rethinks the traditional communications system design. The receiver system utilizes DBNs, more specifically a greedy layer-wise ML algorithm that is able to automatically learn a new latent representation of the data. The two main functions of the receiver system are -- (1) de-noising the received signal and (2) classifying the signals into their binary representatives. The features extracted by the DBN are considered as the properties of input data and are formed by considering the output layer of the DBN. For the first aspect of the proposed receiver, we train a DBN such that it learns to extract features of the received signal. The trained DBN distinguishes the features of the segmented pre-processed signal and groups them with the same ``clean'' framed training data. Furthermore, DBN is capable of reconstructing the input data based on their reduced, learned representation. After the DBN, the classification part of our system uses the features which can be tuned by back-propagation for classification.

The main contributions of this paper are as follows:
\begin{enumerate}
    \item Developed a de-noising DBN model. The trained DBN model distinguishes the features extracted from a segmented pre-processed signal. It then groups these features with the same ``clean'' framed training data.
    \item Redesigned the demodulation block using a classifying DBN model that utilizes feature extraction and back-propagation for classification of the received signals. 
    \item The simulation results show that our proposed DBN based system is able to remain relatively resolute against the different characteristics of the UWAC channel. Furthermore, a sea trial was conducted to verify the performance of the proposed DBN-based receiver system in a real life environment.
\end{enumerate}

The remainder of this paper is organized as follows. Section~\ref{S1} describes the communication system, underwater channel, and receiver system models. Section~\ref{S2} illustrates the proposed DBN based de-noising technique used. Section~\ref{S3} provided a description of the proposed DBN based demodulation technique used in the novel receiver system. Section~\ref{S4} discussed the results of the proposed receiver system and the sea trial used for validation. Finally, conclusions are drawn in Section~\ref{S5}.

\section{System Model Overview}
\label{S1}

\begin{figure*}[h]
\centering
\includegraphics[width=15cm]{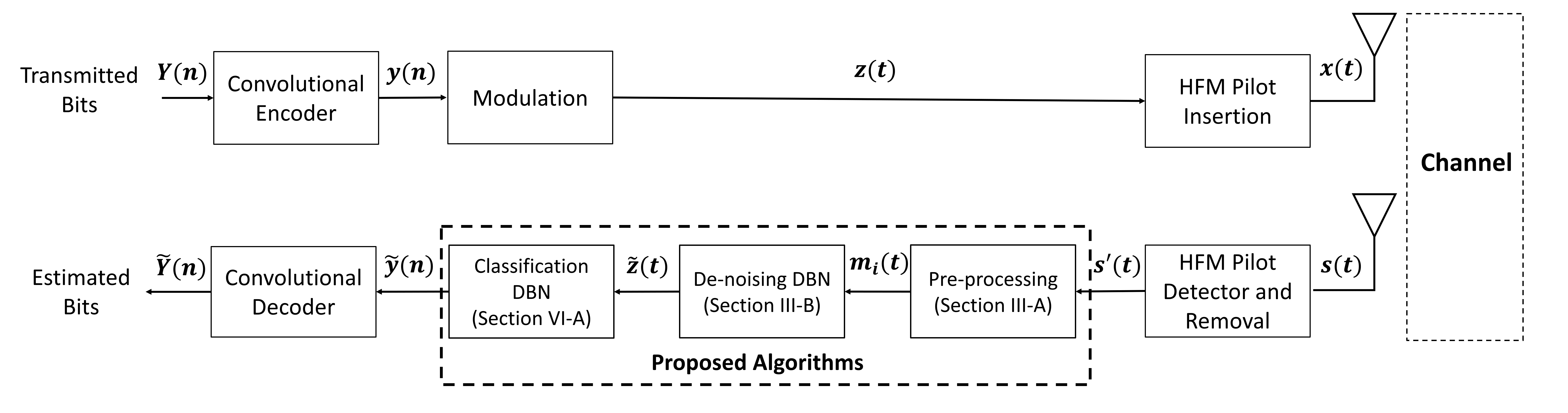}
\caption{End-to-end System Model}
\label{End-to-end System Model}
\end{figure*}

In this section, we describe the proposed end-to-end communication system, represented in Fig. \ref{End-to-end System Model}, comprising of a single transmitter and receiver. 

The following subsections will describe the overall communication system model used to test our proposed system, the derivation of the underwater acoustic channel model, and an overview of the proposed receiver system.

\subsection{Communication System Model}

First, let $y(n)$ be the representation of the convoluted transmitted bits, $Y(n)$ , and the binary representation of the modulated target transmitted signal $z(t)$ during the \textit{n}-th transmission. A series of transmission symbols $y(n)$ are translated into different transmission signal waveforms $z(t)$ via a Phase Shift Key (PSK) modulator as described in \cite{F.Gardner1986}. 

Second, let $x(t)$ represent the overall transmitted signal consisting of $z(t)$ and a Hyperbolic Frequency Modulated (HFM) pilot \cite{M.Kim2017}. $x(t)$ is then relayed through a channel model to obtain the received signal $s(t)$:
\begin{equation}
s(t) = H(t)\cdot x(t) + no(t)
\end{equation}
where $H(t)$ is the channel equation, $\cdot$ represents the dot product, and $no(t)$ is the Additive White Gaussian Noise (AWGN).

Finally, we generate a set of training data $[z_t, y_t]$, $t = 1, 2, ..., n$, where $z_t$ is a training signal, $y_t$ is the corresponding BPSK binary label vector consisting of 0s and 1s, and $n$ is the number of the training signals. Once detection and removal of the HFM pilot is completed, the desired section of the received signal $s'(t)$ acts as the input for the proposed receiver system. The algorithm will build a model from the training data $L$, such that for a given $s'(t)$, the trained model will be able to predict a reconstructed waveform $\Tilde{z}(n)$ and its corresponding label $\Tilde{y}(n)$. Therefore, predicting the received bits, $\Tilde{Y}(n)$. 

\subsection{Underwater Channel}
\label{Underwater Channel}
Underwater acoustic communication channels are regarded by researchers as one of the most complex communication channels to model. 
Multi-path propagation and Doppler effects are recognized as one of the most challenging factors of the underwater acoustic channel \cite{K.C.H.Blom2016,M.Johnson1997,M.Stojanovic1993}. The more common techniques to approximately simulate the underwater acoustic channel vary from signal-to-noise ratio (SNR)-based channel models that rely on empirical equations as seen in \cite{M.Stojanovic2009} to models that are based on the assumption of Rayleigh signal fading in \cite{J.W.Chavhan2015,M.A.2005}.

In this subsection, the channel model used for the simulation will be presented, taking into consideration multi-path propagation and Doppler effect. 

\subsubsection{Multi-path propagation}

Multi-path propagation in the ocean is mostly governed by sound reflecting off underwater surfaces like bubbles and the seabed \cite{T.C.Yang2011}. These effects will cause an elongation of the path traveled, and thus a time delay. The received signal in a mutli-path environment can be generally represented as seen in \cite{A.K.Morozov2006,L.M.Wolff2012}:
\begin{equation}
    s(t) = \sum_{i=0}^{N} A(t) \cdot x(t-\tau_i)
    \label{Multi-path Eq}
\end{equation}
where $A$ represents the reverberation created by the reflection and scattering. This phenomena results in a reflection phase change and a propagation loss \cite{T.C.Yang2011}. As such, Eq. \ref{Multi-path Eq} can be further expressed as:
\begin{equation}
    s(t) = \sum_{i=0}^{N} a^{a}_{i} {a^{b}_i(\theta(t)) \circ x(t - \tau_{i})} + no(t)
    \label{multi-path eq}
\end{equation}
where $a^{a}_i$ is the amplitude variation caused by the reverberation and fading of the channel, $\circ$ represents the Hadamard product, and $a^{b}_i(\theta(t))$ is the phase variation and is modeled as seen in \cite{Liu2015}:
\begin{equation*}
a^{b}(\theta(t)) = [1\quad e^{-j\theta_{1}(t)}...\quad e^{-j\theta_{k-1}(t)} ]
\end{equation*}
where $j$ represents the imaginary number, $k$ is the length of the signal and $\theta_{k}$ is the phase shift corresponding to the change in angle.

\subsubsection{Doppler Effect}
In underwater communications, a combination of the low speed of underwater sound propagation and the relative movement of the transmitter and receiver introduces the Doppler effect \cite{M.Stojanovic2009}. Let $v$ denote the speed of the relative movement of the transmitter and the receiver, and $f_c$ denote the carrier frequency of the transmitted signal. The carrier frequency at the receiver is given by:
\begin{equation}
    f = (1+\frac{\Delta_{rt}}{\Delta_s}) f_c
\end{equation}
where $\Delta_s$ denotes the sound propagation speed underwater. Note that $\Delta_{rt}$ is positive in the event that the receiver is moving toward the transmitter, otherwise $\Delta_{rt}$ is negative. 

In the time domain, the Doppler effect can be construed as a lengthening or compression of the transmitted waveform \cite{G.Eynard2008,Pan2015}. The Doppler effect can be depicted in the time-domain as:
\begin{equation}
     s(t) = x((1-\alpha)t) +no(t)
     \label{Doppler Eq}
\end{equation}
where $\alpha$ is the Doppler co-efficient.

Taking into consideration the above contributing characteristics, the channel model used in this paper is:
\begin{equation}
s(t) = \sum_{i=0}^{N} a^{a}_{i} {a^{b}_i(\theta(t)) \circ x((1-\alpha_i)t - \tau_{i})} + no(t)
\label{Simulation Equation}
\end{equation}

\subsection{Receiver System Model}
\label{Receiver System Model}
The receiver model is comprised of two blocks -- (1) de-noising and (2) demodulation. 

In the de-noising component, the input signal $s'(t)$ is first converted and normalized into a pixelized matrix $m(t)$ via the proposed pre-processing method. $m(t)$ is then partitioned into $i$ number of $m_i(t)$ to meet the requirement of the proposed algorithm for feature extraction. The learning features of the training data is used to find the closest match to the features of $m_i(t)$, which is then used as a basis for reconstruction via:
\begin{equation}
\Tilde{z}(t) = \Psi(W\cdot m_i(t) + b) 
\end{equation}
where $\Psi(\cdot)$ is a learning function, $W$ and $b$ represents the weights and bias of the network. 

In the demodulation block, the reconstructed waveform $\Tilde{z}$ is classified to a label $\Tilde{y(n)}$ via:
\begin{equation}
\Tilde{y}(n) = \Phi(\Tilde{z}(t))
\end{equation}
where $\Phi(\cdot)$ is a learning function. 

The focus of this paper is then to optimize the learning functions, $\Psi(\cdot)$ and $\Phi(\cdot)$, and their corresponding weights and bias.

\section{Proposed DBN-based De-noising}
\label{S2}
\begin{figure}[h]
\centering
\includegraphics[width=8.5cm]{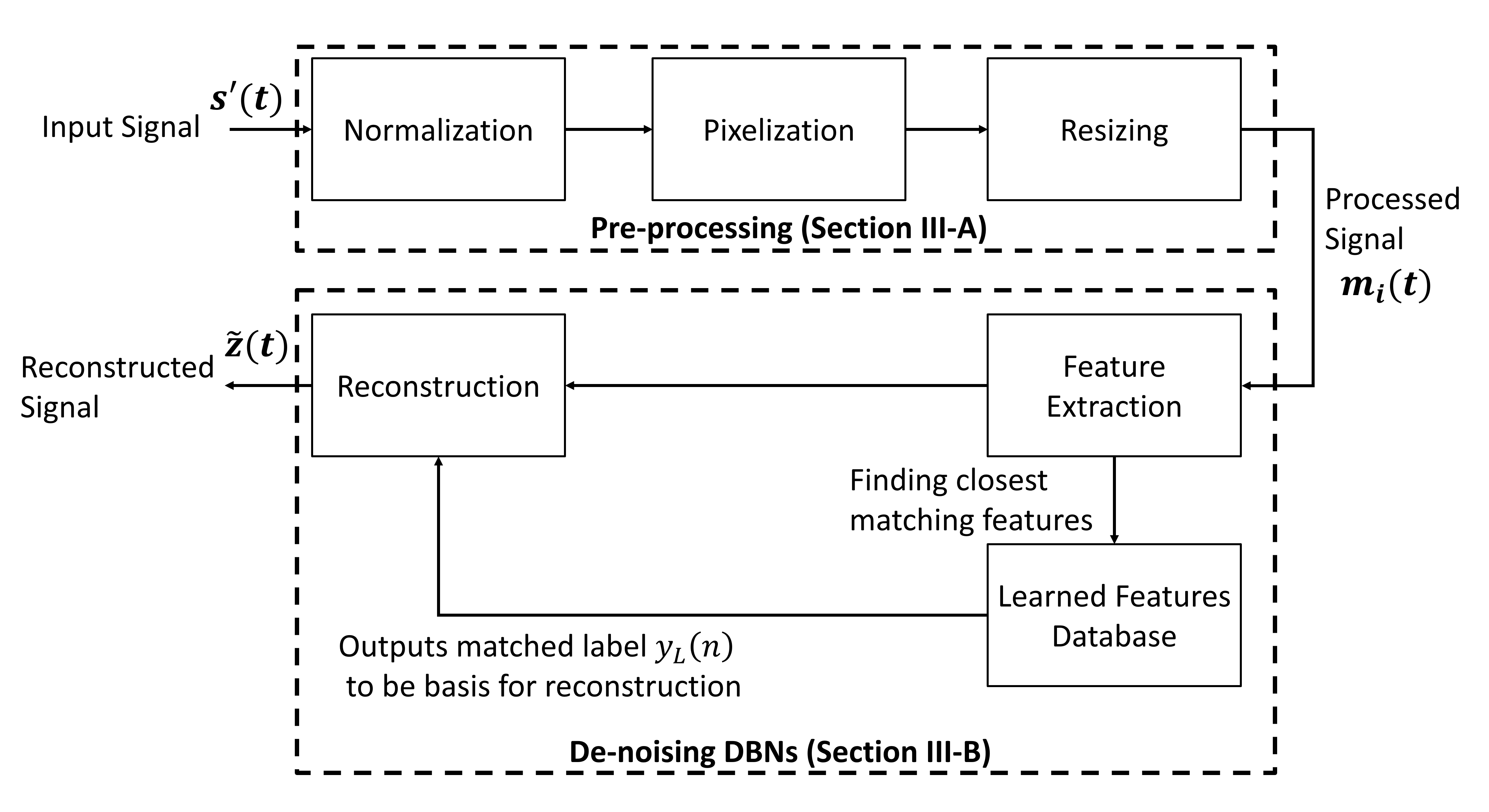}
\caption{Overview of De-noising Block}
\label{Overview of De-noising Block}
\end{figure}

In this section, the de-noising algorithm, consisting of both the pre-processing method and the de-noising DBN, is described. An overview of our proposed algorithm is shown in Fig.\ref{Overview of De-noising Block}. The input is the received signal $s'(t)$ and the output is the reconstructed signal $\Tilde{z}(t)$.

\subsection{Pre-processing}

\begin{figure}[h!]
\centering
    \begin{subfigure}[b]{0.45\textwidth}
    \centering
        \includegraphics[width=\textwidth]{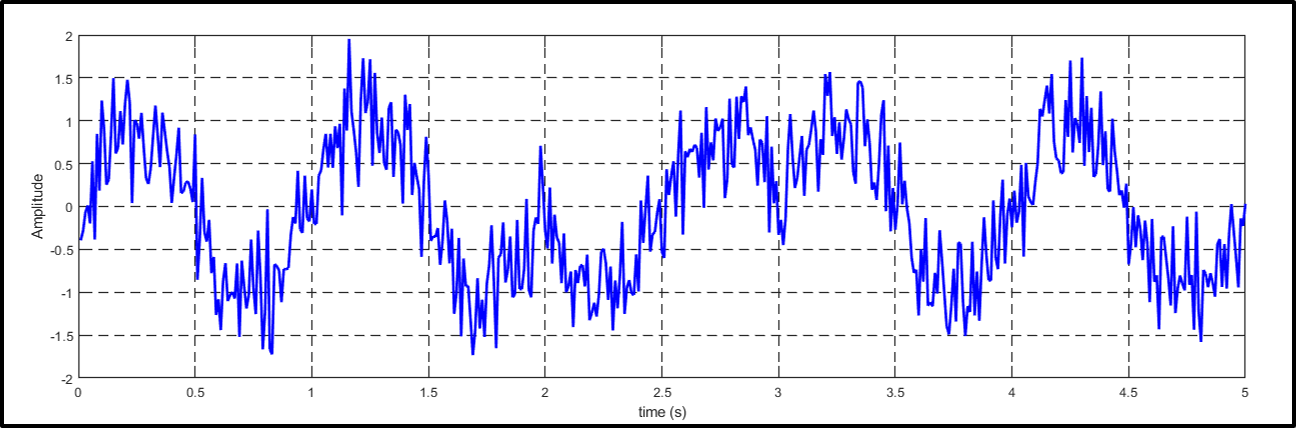}
        \caption{Received Signal $s'(t)=1 \times n$ Matrix}
        \label{Received Signal s(t)}
    \end{subfigure}\\
    \begin{subfigure}[b]{0.45\textwidth}
    \centering
        \includegraphics[width=\textwidth]{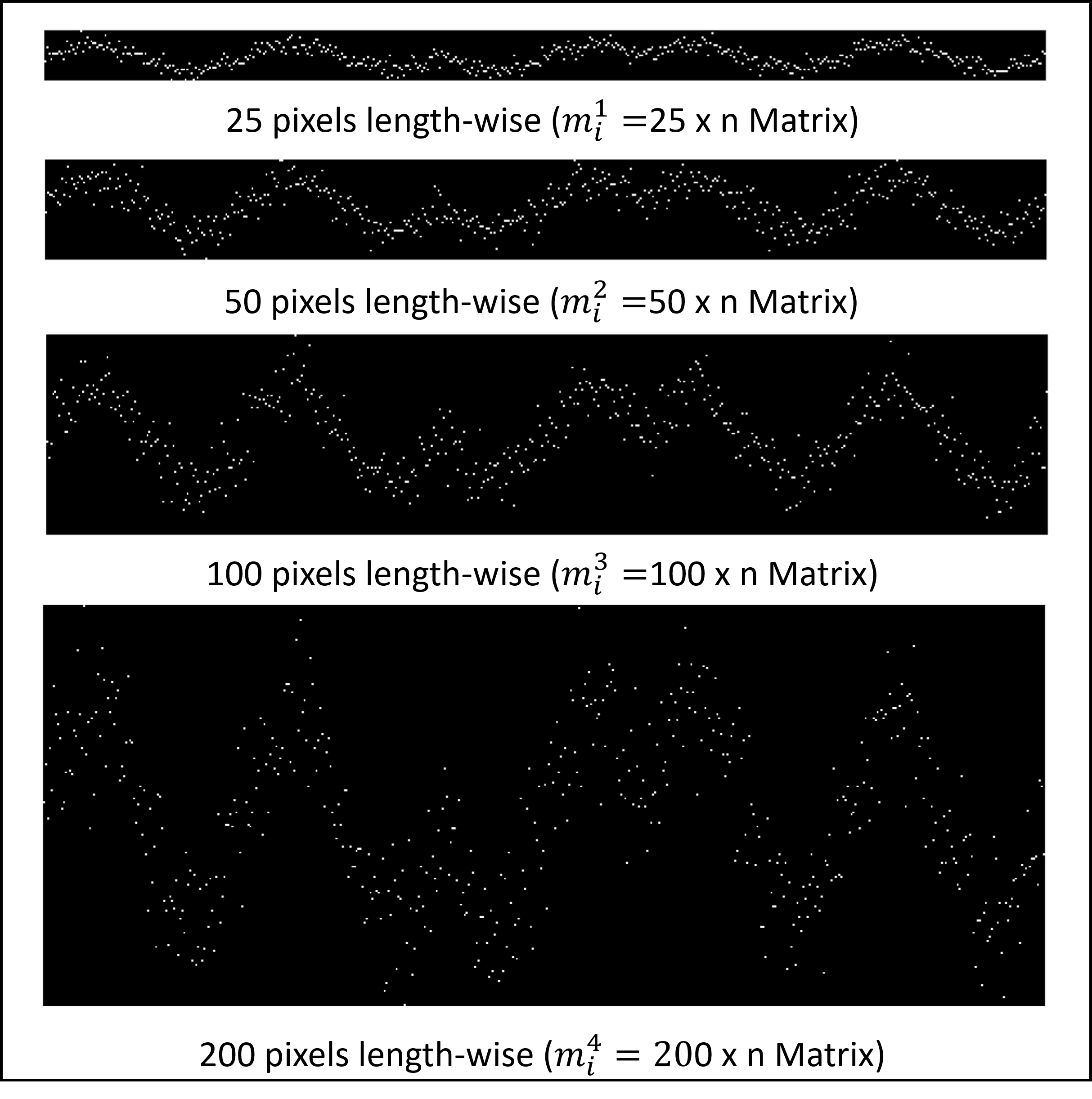}
        \caption{Pre-processed Signal $m_i(t)$}
        \label{Pre-processed Signal}
    \end{subfigure}\\
    \caption{Visualization of the Pre-processing Block Diagram presents the received signal (input) and the pixelized matrix $m_i(t)$ (output). The received signal is pre-processed into 4 matrices to provide the DBN based de-noising algorithm with more features to extract.}
    \label{Visualization of Pre-processing Block Diagram}
\end{figure}

\begin{figure}
\centering
\begin{algorithm}[H]
\caption{Pixelization algorithm}\label{Pixelization Algor}
\begin{algorithmic}[1]
\Require{$s'$ and $Pix$}
\Ensure{$m$}
\Statex
\State $length_s \gets \texttt{length of } s'$
\State $m \gets \texttt{ones}(Pix, F_l);$
\State $Res \gets \frac{1}{Pix};$
\State $range \gets 1:-Res:0;$
\For{$i \gets 1$ to $length_s$}  
    \State {$D$ $\gets$ {$range'-s'(i)$}}
    \State {$A$ $\gets$ {$min(abs(D))$}}
    \State {$loc$ $\gets$ {$D[A]$}}
    \State {$m(loc,i)$ $\gets$ $0$}
\EndFor
\State \Return {$m$}
\end{algorithmic}
\end{algorithm}

\end{figure}

The pixelization method is defined as:
\begin{equation*}
    m(t)=f(\textit{normalized}(s'(t)))
\end{equation*}

The input signal $s'(t)$ is first normalized into the range of~0 to~1. We then proceed to pixelize the signal to form $m(t)$. Let $Pix$ be the number of pixels (length wise), which controls the resolution of the pixelization. The implemented pixelization algorithm is shown in Algorithm \ref{Pixelization Algor}. The input and output of the pixelization is shown in Fig.\ref{Received Signal s(t)} and Fig.\ref{Pre-processed Signal} respectively. 

Lastly, $m(t)$ is resized to various resolutions, as shown in Fig. \ref{Visualization of Pre-processing Block Diagram}. This allows for more features to be extracted and used for the reconstruction. 

\subsection{De-noising DBN (stacked RBMs)}
\label{De-noising DBNs}

\begin{figure}[h]
\centering
\includegraphics[width=8cm]{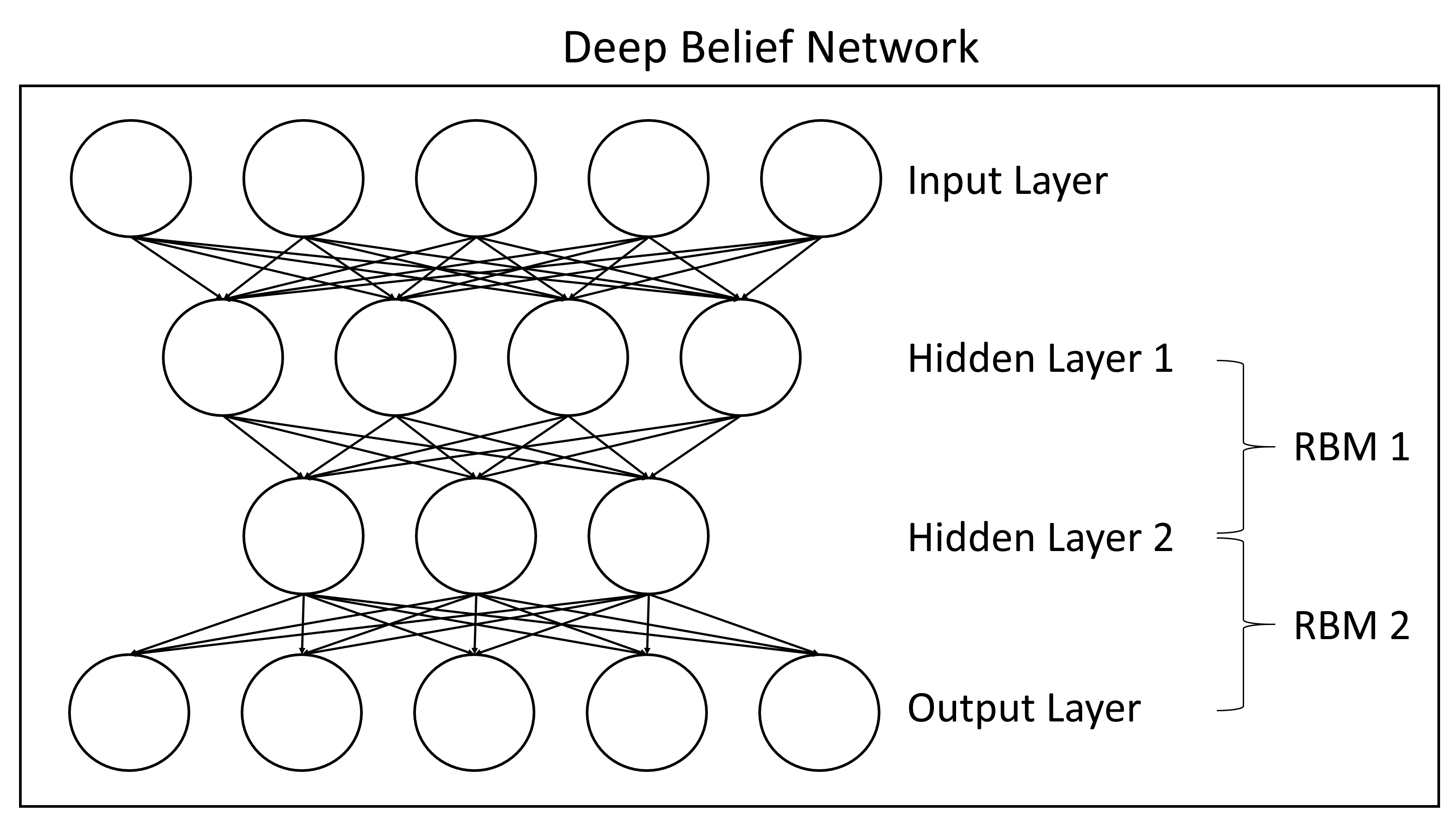}
\caption{Overview of a DBN consisting of stacked RBMs}
\label{Overview of a DBN consisting of stacking RBMs}
\end{figure}

DBNs are probabilistic generative algorithms which provide a joint probability distribution over observable data and labels. Restricted Boltzmann Machines (RBMs) are the building blocks of a DBN. Hence, in this section first we briefly describe RBMs and then we will explore DBN. Fig.\ref{Overview of a DBN consisting of stacking RBMs} illustrates the concept of stacking RBMs to form a DBN. 

A Boltzmann Machine (BM) is a particular form of a Markov Random Field (MRF), where its energy function is linear in its free parameters. Some of its variables (hidden units) allow the machine to represent complicated distributions internally. However, they are unobserved.

\subsubsection{RBMs}
The energy function of the joint configuration in Boltzmann machines is given as follows:
\begin{equation}
\begin{split}
    E(v,h) &= -\sum_{k=1}^{u_v} \sum_{j=1}^{u_h} hWv-\sum_{k=1}^{u_v}b v-\sum_{j=1}^{u_h}c h \\
    &=\mathbf{-h^{T}Wv-b^{T}v-c^Th}
    \label{Energy Function}
\end{split}
\end{equation}
where the visible nodes $v \in \mathbb{R}$ correspond to the input and $u_v$ is the number of visible nodes, the hidden nodes $h \in \mathbb{R}$ represent the latent features and $u_h$ is the number of hidden nodes, $W$ represents the concurrent weights linking the nodes of the visible to the hidden layer, $b$ and $c$ are the bias terms of the hidden and visible nodes respectively.

The free energy can also be expressed in the following form:
\begin{equation}
\begin{split}
    F(v) &=-\log{\sum_{\mathbf{h}} e^{-E(v,h)}}\\
    &=-\log{(\mathbf{h^T W v}+\mathbf{b^T v}+\mathbf{c^Th})}\\
    &=-\mathbf{b^T v}-\sum_{j=1}^{u_h} \log \sum_{h_j} e^{h_j(c_j+\mathbf{W_j v})}
    \label{Free Energy}
\end{split}
\end{equation}

Because visible and hidden units are conditionally independent of one-another, the following equations hold true. 
\begin{align}
    P(v|h) = \prod_{j=1} p(v_j|h)\\
    P(h|v) = \prod_{k=1} p(h_k|v)
\end{align}

When binary units are used, so that $v_j$ and $h_k \in {0, 1}$, and a probabilistic version of the usual neural activation is obtained:
\begin{align}
    P(v_j=1|h) &= \mathit{sigm} (b_j+\mathbf{W_j^T h})\\
    P(h_k=1|v) &= \mathit{sigm} (c_k+\mathbf{W_k v})
\end{align}

The free energy of an RBM with binary units becomes
\begin{align}
    F(v) &=-\mathbf{b^T v}-\sum_{j=1}^{u_h} \log(1+e^{h_j(c_j+\mathbf{W_j v})})
    \label{Binary Free Energy}
\end{align}

Since RBMs are energy based algorithms, i.e. they associate a scalar energy to each configuration of the variables of interest, training them corresponds to modifying that energy function so that its shape has desirable properties, such as low energy configurations.

Energy-based probabilistic models define a probability distribution through an energy function, as follows:
\begin{equation}
    P(v,h) = \frac{1}{Z} \exp{(-E(v,h))}
    \label{Energy partition}
\end{equation}
where $Q$ is the partition function that is obtained via:
\begin{equation}
    Q = \sum_{k=1}^{u_v} \sum_{j=1}^{u_h} \exp{(-E(v,h))}
    \label{Energy partition}
\end{equation}

To optimize the parameters of the network at each layer $k$, the following optimization problem shown by Eqn.~\ref{Objective} is minimize via partial differentiation with respects to $W,b,c$.
\begin{equation}
    g_k(v,h) = - \frac{1}{m} \sum_{j=1}^{m}\log(P(v_k^j,h_k^j))
    \label{Objective}
\end{equation}

\subsubsection{Stacking RBMs into a DBN}
A DBN is comprised of stacked restricted Boltzmann machines with a fast-learning algorithm that allows the structure to achieve better results with less computational effort. 

It models the joint distribution between an observed vector $x$ and $l$ hidden layers $h_k$ as follows:

\begin{equation}
\begin{split}
     P(\mathbf{x},\mathbf{h_1},...,\mathbf{h_l})&=\Bigg[\prod_{k=0}^{l-2} P(\mathbf{h_k}|\mathbf{h_{k+1}})\Bigg] P(\mathbf{h_{l-1}}|\mathbf{h_l})
 \end{split}
\end{equation}
   
where $\mathbf{x}=\mathbf{h_0}$, $P(\mathbf{h_{k}}|\mathbf{h_{k+1})})$ is a conditional distribution for the visible units conditioned on the hidden units of the RBM at level $k$, and $P(\mathbf{h_{l-1}}|\mathbf{h_{l}})$ is the visible-hidden joint distribution (output).

\begin{table}[h]
\centering
\caption{The de-noising DBN structure consist of 2 RBMs.}
\label{De-noising DBN consist of 4 RDMs}
\scalebox{1}{%
\begin{tabular}{c|lc|c}
\toprule
Input & \multicolumn{2}{c|}{Structure} & Output \\ \midrule
\multirow{4}{*}{$m_i(t)$} & No of layers & 2 & \multirow{4}{*}{$\Tilde{z}(t)$} \\
 & Nodes & {[}875,625{]} &  \\
 & Activation Function & {[}Sig, Sig{]} &  \\
 & Epoch & 1000 & \\ \bottomrule
\end{tabular}
}
\end{table}

\subsubsection{Training}

To train a DBN such that it can perform matrix de-noising, the normalized pixel values of the pixelized signal are used as input. By using min-max normalization, $m_i(t)$ is transformed into a floating-point number system with a range of 0 and 1. Unlike the first and last layer of the DBN, hidden layers consist of binary nodes. The main idea is to train a DBN to be able to associate noisy $m_i(t)$ to $m_i(t)$ with lower noise or no noise. This idea can be implemented by learning the features extracted from the noisy and clean $m_i(t)$ contents. These features are then presented in some nodes at the last layer of the network. 

The network is trained with a variety of noisy $m_i(t)$ as input and clean $m_i(t)$ as the desired output. Using a standard basis called relative activity to detect noise nodes, each node is defined as the difference between two values of a particular node which results from feeding the network a clean $m_i(t)$ and its corresponding noisy $m_i(t)$. As a result, if a particular node is a noise node, it should have higher relative activity. On the other hand, if it is a clean noiseless node, it should have a lower relative activity. This theory is justified by the fact that the activation of $m_i(t)$ nodes should be same for both clean noiseless and its corresponding noisy $m_i(t)$.

By performing the above action for all $m_i(t)$ and averaging the values of the last layer's nodes, the average relative activity of the last layer is computed. The nodes with a higher average relative activity are still viewed as noise nodes. Once the noise nodes are discovered, the next step is to lower their activity by selecting the average value of all the noise nodes as their neutral values. As such, the noise nodes are then considered inactive and a clean noiseless $m_i(t)$ can be reconstructed.

\subsection{Results of DBN based De-noising}
In this subsection, we evaluate the proposed DBN based de-noising technique. As a baseline for comparison, we used the conventional MLE method devised in~\cite{Stocia1989} and the de-noising auto encoder in~\cite{Goodfellow}. To analyze the only performance of the de-noising capability, the system used was uncoded.

For the following simulation experiments, the simulated BPSK dataset contains 100,000 transmitted signals periods, in which 50\% is used for training, 20\% on validation and the remaining 30\% on testing. The dataset was generated using Eq.\ref{Simulation Equation} and Table.\ref{mean}. The $f_c$, sampling frequency $f_s$ and bit rate $Rb$ of the BPSK signals were set at $2kHz$, $40kHz$ and $1kbits/s$. The frequency of random change, $f_\delta$, was 2kHz. For consistency, the de-noising auto encoder used as a comparison in this section was trained using the same dataset. 

\begin{table}[h]
\centering
\caption{Mean and Standard deviation of random distribution for simulated channel parameters}
\label{mean}
\scalebox{1}{%
\begin{tabular}{c c c}
\toprule
Parameter & $\mu$ & $\sigma$\\ \midrule
$a^{a}$ & 0.75 & 0.25\\ 
$a^{b}(\theta)$ & $\pi$ & $\frac{\pi}{2}$\\
$\tau$ & $\frac{fs}{2}$ & $\frac{fs^2}{2}$\\ \bottomrule
\end{tabular}}
\end{table}

\begin{figure}[h]
\centering
\includegraphics[width=8.5cm]{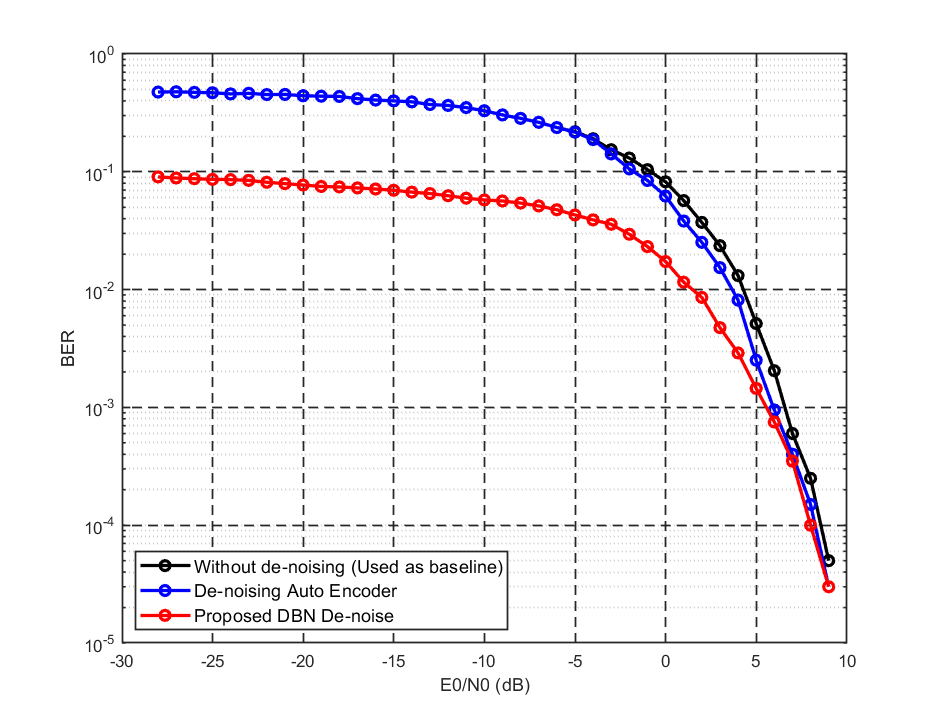}
\caption{De-noising BER Accuracy comparison with uncoded BPSK under AWGN channel. For consistency, MLE was used as the demodulation technique for all 3 methods.}
\label{De-noising Demodulation Accuracy comparison with BPSK (AWGN)}
\end{figure}

First, we conducted a simulation experiment to evaluate the proposed DBN based de-noising technique's ability to remove noise for channels with extremely high noise. Fig.\ref{De-noising Demodulation Accuracy comparison with BPSK (AWGN)} shows the Bit Error Rate (BER) of the proposed DBN based de-noising technique under the AWGN channel. As a baseline for comparison, we have provided the BER of the MLE and de-noising auto encoder to highlight the substantial gains for highly negative $\frac{Eb}{No}$. A reason for this could be the existence of noise invariable properties in the features extracted by the DBN in the proposed DBN based de-noising technique. Evidence of this can be seen by the converging performance of the algorithm to the baseline as the noise level decreases, resulting in a decrease in functionality of the noise invariant property. At BER of $10^{-2}$, the performance of the proposed DBN based de-noising technique has a significantly smaller gain of 2.4dB for de-noising auto encoder and 2.6dB for the MLE. 

\begin{figure}[h]
\centering
\includegraphics[width=8.5cm]{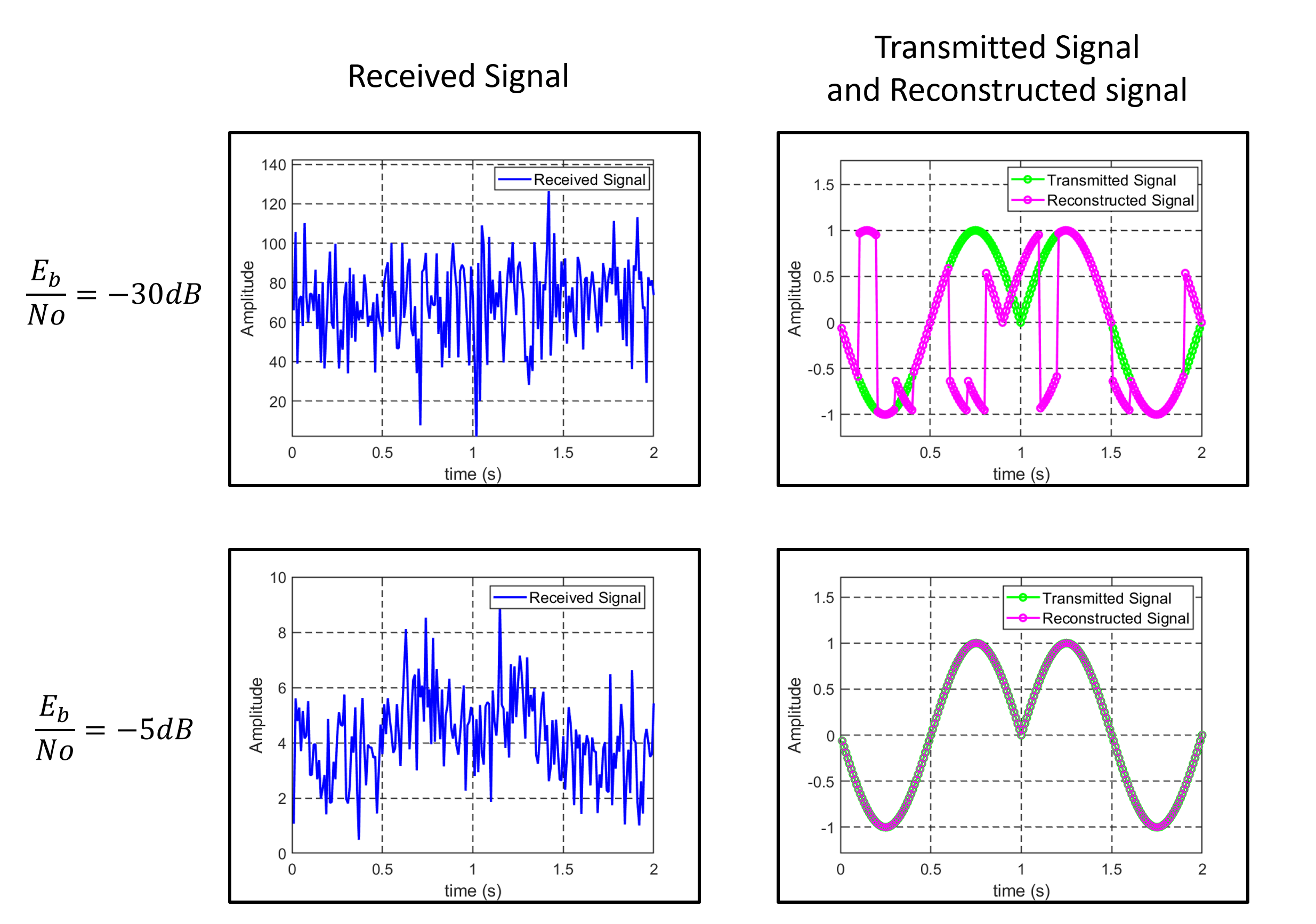}
\caption{Visualization of proposed DBN based de-noising algorithm under AWGN channel presents the received signal (input) and the reconstructed signal (output). The reconstructed signal, depicted by the magenta line, is shown in relation to the transmitted signal (ideal output), depicted in green.}  
\label{Reconstruction of Received Signal}
\end{figure}

Fig.\ref{Reconstruction of Received Signal} shows a visualization of the de-noising outcome. At $\frac{Eb}{No}=$-30dB, the received signal is highly distorted by the channel noise. However, the proposed technique is still able to partially predict the waveform shape. At $\frac{Eb}{No}=$-5dB, the waveform can be almost perfectly reconstructed.

\begin{figure}[h]
\centering
\includegraphics[width=8.5cm]{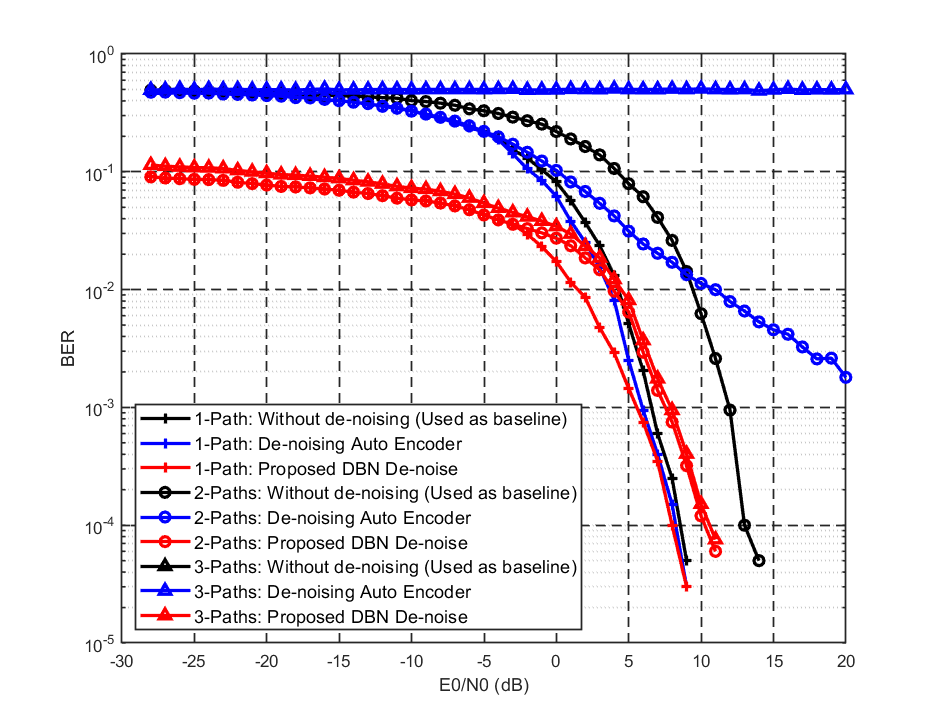}
\caption{De-noising BER Accuracy comparison with uncoded BPSK under Multi-path channel, modelled in Eq.\ref{multi-path eq}. For consistency, MLE was used as the demodulation technique for all 3 methods.}
\label{De-noising Demodulation Accuracy comparison with BPSK(Multi-path channel)}
\end{figure}

The second simulation experiment we conducted tested the algorithm's ability to remain resolute against multi-path propagation. The simulated channel distorted received signals, utilized as test cases in this experiment, were modelled by Eq. \ref{Multi-path Eq}. The number of multi-paths in the dataset was distributed as 40\% 1-path, 30\% 2-paths and 30\% 3-paths. The results of the experiment are shown in Fig.\ref{De-noising Demodulation Accuracy comparison with BPSK(Multi-path channel)}. With the increasing number of paths, the BER of the proposed DBN based de-noising algorithm achieves considerable gains while remaining relatively stable in comparison to the de-noising auto encoder and MLE.

\begin{figure}[h]
\centering
\includegraphics[width=8.5cm]{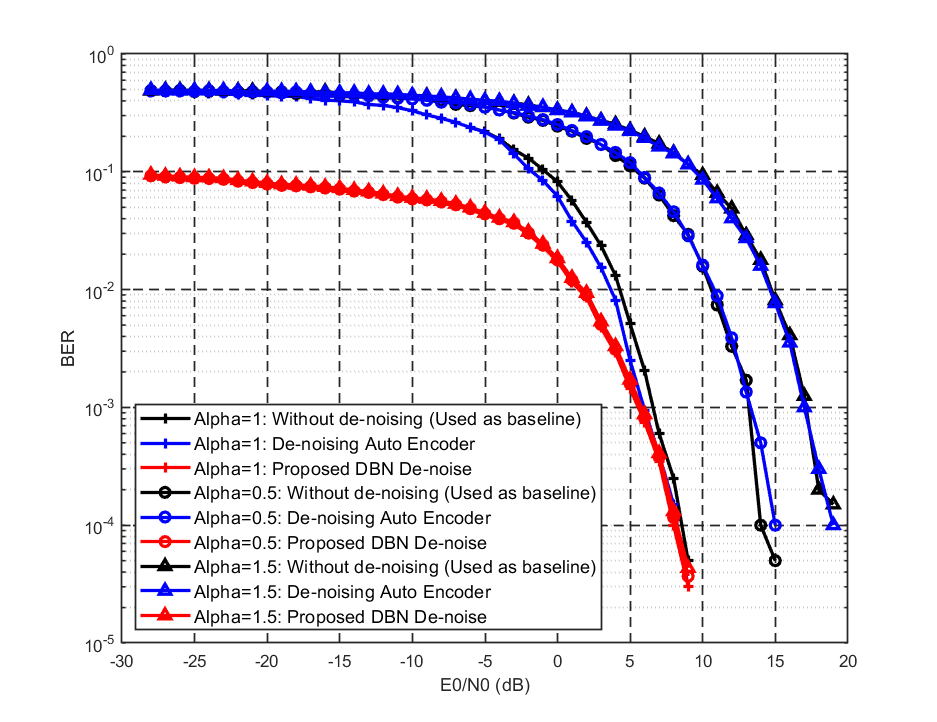}
\caption{De-noising Demodulation BER Accuracy comparison with uncoded BPSK, modelled in Eq.\ref{Doppler Eq}. For consistency, MLE was used as the demodulation technique for all 3 methods.}
\label{De-noising Demodulation Accuracy comparison with BPSK (Doppler channel)}
\end{figure}

Finally, to test the influence of the Doppler effect on the proposed DBN based de-noising algorithm, a simulation experiment was conducted using Eq.\ref{Doppler Eq} for the channel. Fig.\ref{De-noising Demodulation Accuracy comparison with BPSK (Doppler channel)} depicts the results under three different scenarios, where the $\alpha=0.5,1,1.5$ resulting in a $f_c= 1kHz,2kHz,3kHz$. The BER of the proposed algorithm for all three scenarios are observed to be similar. Thus implying that for a certain range of $\alpha$, the algorithm is able to remain relatively rigid to the influences of the Doppler effect.

\section{Proposed DBN-based Demodulation}
\label{S3}
In this section, the demodulation algorithm, consisting of a classification DBN, is described. 

\subsection{Classification DBN}

For the classification DBN, the same general stacked energy based RBM algorithm is used as described in Section \ref{De-noising DBNs}. The input is the reconstructed signal $\Tilde{z}(t)$ and the output consists of the respective binary labels $\Tilde{y}(n)$ of 0s and 1s. 

\subsection{Determining the structure of Classification DBN}

To determine the classification DBN structure, we investigated the influence of different network structures on the performance of the algorithm in the classification task at $\frac{Eb}{No}=$ 0dB. As shown in Table \ref{tab:my-table}, we attempted nine different network structures, which consist of a varying number of hidden units, and trained them for a varying number of epochs. The best classification BER results obtained was using the $[1250,50]$ structure. Although the structure $[1250,100]$ seems to achieve approximately the same results, the time needed for training is significantly larger.

\begin{table}[h]
\centering
\caption{Effects of training epoch and number of nodes on Classification DBN}
\label{tab:my-table}
\scalebox{1.2}{%
\begin{tabular}{c c c c}
\toprule
\multirow{2}{*}{\begin{tabular}[c]{@{}c@{}}Number of units\\ {[}Layer 1,Layer 2{]}\end{tabular}} & \multicolumn{3}{c}{Results} \\ \cline{2-4} 
 & Epoch Number & BER & Time (s) \\ \midrule
$[650,50]$ & 200 & 0.1712 & 65.79 \\ 
$[650,50]$ & 500 & 0.1386 & 378.2\\ 
$[650,50]$ & 1000 & 0.1121 & 2671 \\ 
$[1250,50]$ & 200 & 0.0854 & 1261\\ 
$[1250,50]$ & 500 & 0.0825 & 5731\\ 
$[1250,50]$ & 1000 & 0.0805 & 78106\\ 
$[1250,100]$ & 200 & 0.0883 & 5228\\ 
$[1250,100]$ & 500 & 0.0844 & 18674\\ 
$[1250,100]$ & 1000 & 0.0810 & 162760\\ \bottomrule
\end{tabular}}
\end{table}

To minimize complexity and maximize the performance of the algorithm, we have chosen to use the structure as illustrated in Table \ref{Classification DBN consist of 2 RDMs}.

\begin{table}[h]
\centering
\caption{The final hyperparameters used in our proposed classification DBN, which consists of 2 layers of RDMs.}
\label{Classification DBN consist of 2 RDMs}
\scalebox{1}{%
\begin{tabular}{c|lc|c}
\toprule
Input & \multicolumn{2}{c|}{Structure} & Output \\ \midrule
\multirow{4}{*}{$\Tilde{z}(t)$} & No of layers & 2 & \multirow{4}{*}{$\Tilde{y}(t)$} \\
 & Nodes & {[}1250,50{]} &  \\
 & Activation Function & {[}Sig, Sig{]} &  \\
 & Epoch & 1000 & \\ \bottomrule
\end{tabular}
}
\end{table}

\subsection{Results of Classification DBN}
In this subsection, we evaluate the proposed classification DBN. As a baseline for comparison, we used the conventional MLE method devised in~\cite{Stocia1989} to illustrate the similar performance of the demodulation techniques.

\begin{figure}[h!]
\centering
\includegraphics[width=8.5cm]{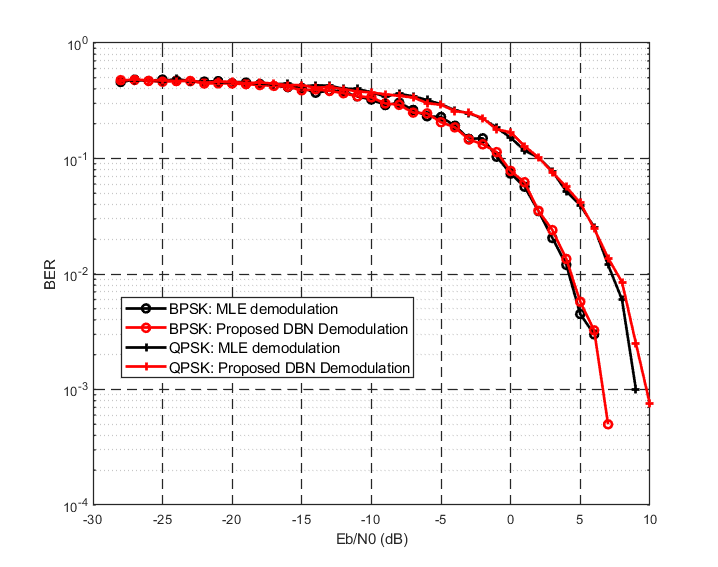}
\caption{Demodulation BER Accuracy comparison with BPSK and QPSK}
\label{Demodulation accuracy comparison of with BPSK and QPSK}
\end{figure}

For the following simulation experiments, the simulated dataset contains 100,000 transmitted signals periods, in which 50\% is used for training, 20\% for validation, and the remaining 30\% for testing. The dataset was generated using a AWGN channel model at a range of $\frac{Eb}{No}=$ -10dB to 30dB. For a fair comparison with MLE, the $f_c$ of the BPSK signals is set at 2kHz. 

Using the MLE as a baseline, this experiment illustrates that the demodulation performance level of the proposed classification DBN is similar to MLE. The results are shown in Fig.~\ref{Demodulation accuracy comparison of with BPSK and QPSK}. This implies that the classification DBN has learned to extract significant features from the PSK modulation scheme. For a truly fair comparison, the proposed algorithm is also compared to Quadrature Phase Shift Keying (QPSK), derived in \cite{F.Gardner1986}, without much extra training. As seen, at BER $10^{-3}$, the algorithm's performance for QPSK has a BER of 0.67dB less in comparison to MLE. A more inclusive training dataset for higher-order modulation schemes could increase the performance of the algorithm in this area. 

\section{Results and Discussion}
\label{S4}

This section will evaluate the proposed receiver as a whole as seen in Fig.\ref{End-to-end System Model}. First, the performance of the receiver will analyzed using the simulated underwater model shown in Eq.\ref{Simulation Equation}. Then, the conditions of the conducted sea trial will be described. Finally, the collected sea trial data was used to validate the real application of the proposed receiver system.

\begin{figure}[h]
\centering
\includegraphics[width=8.5cm]{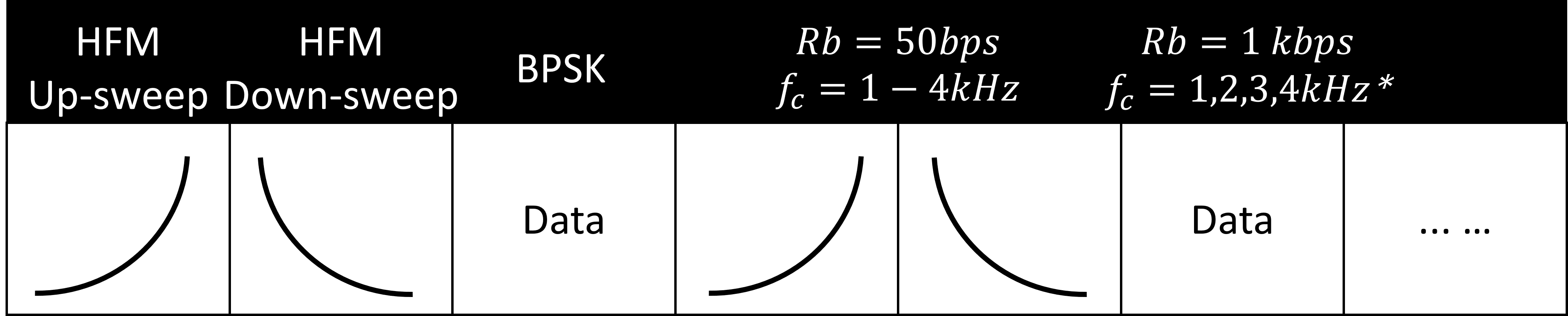}
\caption{Transmitted Data Structure}
\label{Data Structure}
\end{figure}

The data frame of the testing dataset used in both the simulation experiments and sea trials is shown in Fig.\ref{Data Structure}. The pilot consists of a single up-sweep and a down-sweep HFM signal, which is used for detection of the incoming received data signal. The HFM modulated signal has a bit rate of 50 bits/s and a frequency range of 1-4kHz. The data frame includes 416 bits of coded BPSK modulated signals. The specifications of the data structure is recorded in Table.\ref{Specifications on the experimental data set}. 

\begin{table}[h]
\centering
\caption{Specifications on the simulated and sea trial data set}
\label{Specifications on the experimental data set}
\scalebox{1.1}{%
\begin{tabular}{l c}
\toprule
Experiment Parameters & Value\\ \toprule
Sampling Rate &  40 kHz \\ \midrule
Bit Rate of HFM &  50 bits/s \\ \midrule
$f_c$ of HFM & 1 -- 4 kHz\\ \midrule
Bit Rate of BPSK &  1 kbits/s \\ \midrule
$f_c$* of BPSK & 1, 2, 3 \& 4 kHz\\ \bottomrule
\end{tabular}}
\end{table}

\subsection{Simulation Overall Results}

In this subsection, we evaluate the overall proposed receiver system. To assess performance under a underwater environment, we will be employing 5 systems for evaluation -- (1) MLE demodulation, (2) de-noising auto encoder with MLE demodulation, (3) the proposed DBN based receiver, (4) DL orthogonal frequency-division multiplexing (OFDM) \cite{Youwen2019}, and (5) SIC DL \cite{Wang2019IEEE}.

For the following simulation experiments, the training data and channel model used to train the individual parts of the proposed receiver system were the same as stated in Section \ref{S3} and \ref{S4}. For equitable contrast, the de-noising auto encoder did not go through any extra training.

The simulated BPSK testing dataset contains 10,000 transmitted signals periods. The dataset was generated using Eq.\ref{Simulation Equation} and the random distributions seen in Table.\ref{mean1}. The number of multi-paths in the dataset was distributed as 40\% 1-path, 30\% 2-paths and 30\% 3-paths. The dataset contains dataset of 60\% $1kHz$ $f_\delta$ and 40\% $2kHz$ $f_\delta$ in each multi-path cluster. The increase in $f_\delta$ is used to further simulate the complex occurrence of the underwater scattering. To fairly evaluate the performance of the proposed receiver with the two systems mentioned above, the $f_c$ of the BPSK signals is set at 2kHz. 

\begin{table}[h]
\centering
\caption{Mean and Standard deviation of random distribution for simulated overall channel parameters}
\label{mean1}
\scalebox{1}{%
\begin{tabular}{c c c}
\toprule
Parameter & $\mu$ & $\sigma$\\ \midrule
$a^{a}$ & 0.75 & 0.25\\ 
$a^{b}(\theta)$ & $\pi$ & $\frac{\pi}{2}$\\
$\alpha$ & $1$ & $0.5$\\
$\tau$ & $\frac{fs}{2}$ & $\frac{fs^2}{2}$\\ \bottomrule
\end{tabular}}
\end{table}

In a previous investigation seen in \cite{Lee}, we discovered that the feature extraction ability of the DBN has created a characteristic that is invariant to the influences of the Doppler effect. Therefore, we assume that even though the classification DBN was only trained on $f_c=2kHz$, the performance of the proposed classification DBN will not be significantly degraded by the range of $f_c$ used in the testing dataset used. 

\begin{figure}[h!]
\centering
\includegraphics[width=8.5cm]{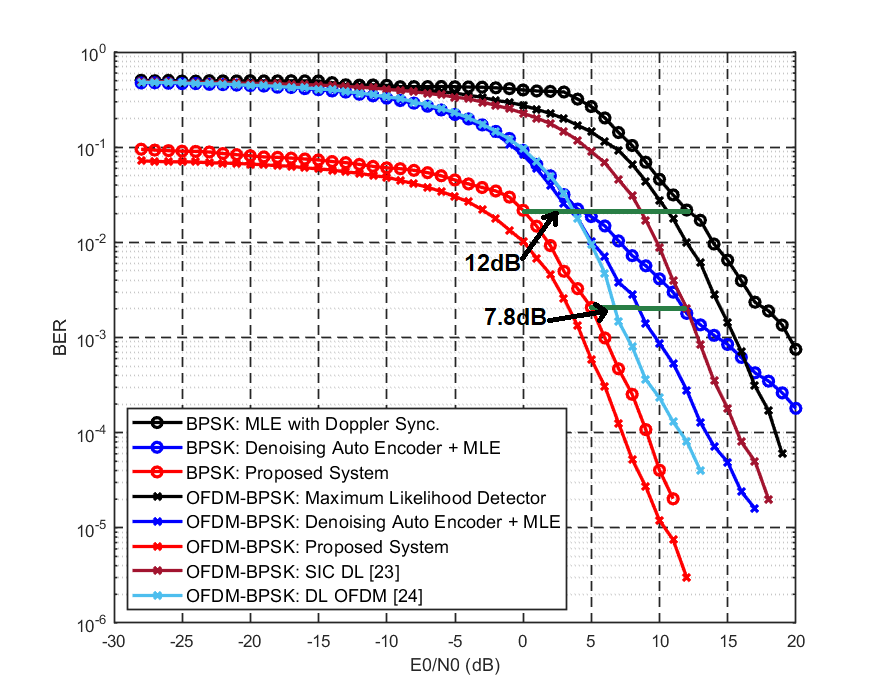}
\caption{BER Accuracy comparison with coded BPSK and OFDM-BPSK under simulated underwater conditions, modelled by Eq.\ref{Simulation Equation}}
\label{Simulation under sea trial conditions}
\end{figure}

Fig.\ref{Simulation under sea trial conditions} depicts the performance of the five systems with regards to the above described testing scenario. The proposed receiver (consisting of both DBN De-noise and DBN Classification) is seen to outperform the other algorithms over a large range of $\frac{Eb}{No}$ for both uncoded BPSK and OFDM-BPSK. This implies that the proposed receiver system is able to remain invariant to changes in instantaneous amplitude, phase and frequencies, such that the shown coding gain can be achieved. 

Table.\ref{Complexity} compares the computational complexity of the five algorithms, where $n$ represents input size $n$ for each function. The results show that our proposed system requires a large amount of training time in comparison to the auto encoder and MLE. However, shown in Fig.\ref{Simulation under sea trial conditions}, our proposed algorithm outperformed the auto encoder and MLE by 7.8dB and 12dB at $\frac{Eb}{N0}=5$ and $\frac{Eb}{N0}=0$ respectively for the BPSK modulated system.

\begin{table}[h]
\centering
\caption{Algorithmic Computational Complexity Comparison}
\label{Complexity}
\scalebox{1.0}{%
\begin{tabular}{c c c}
\toprule
& Training & Testing\\ \midrule
MLE & - & O($2^n$)\\
Auto-encoder & O(1000$n^2$) & O(100$n^2$)\\ 
DL OFDM \cite{Youwen2019} & O(19200$n^2$) & O(9600$n^2$) \\
DL SIC \cite{Wang2019IEEE} & O(9600$n^2$) & O(4800$n^2$) \\
Proposed System & O(16000$n^2$) & O(100$n^2$) \\ 
\bottomrule
\end{tabular}
}
\end{table}

\subsection{Sea Trial Set-up}
\label{Experimental Set-up}

\begin{figure}[h!]
\centering
    \begin{subfigure}[b]{0.5\textwidth}
    \centering
        \includegraphics[width=8.5cm]{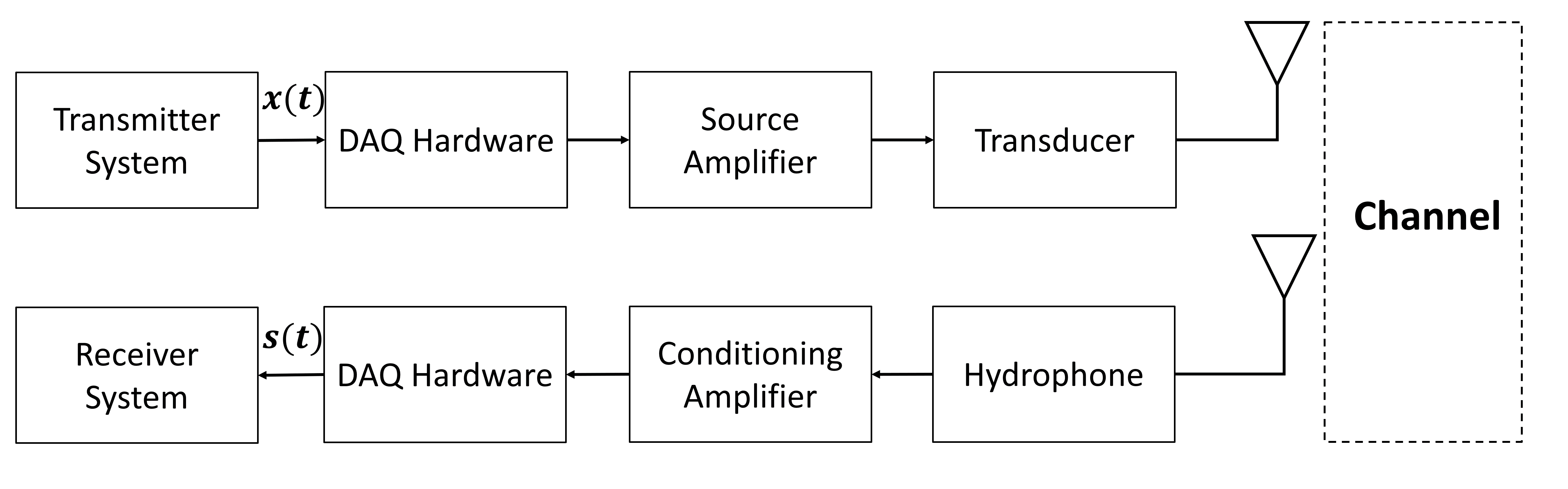}
        \caption{Sea Trial End-to-End System Diagram}
        \label{Sea Trial End-to-End System Diagram}
    \end{subfigure}\\
    \begin{subfigure}[b]{0.4\textwidth}
    \centering
        \includegraphics[width=\textwidth]{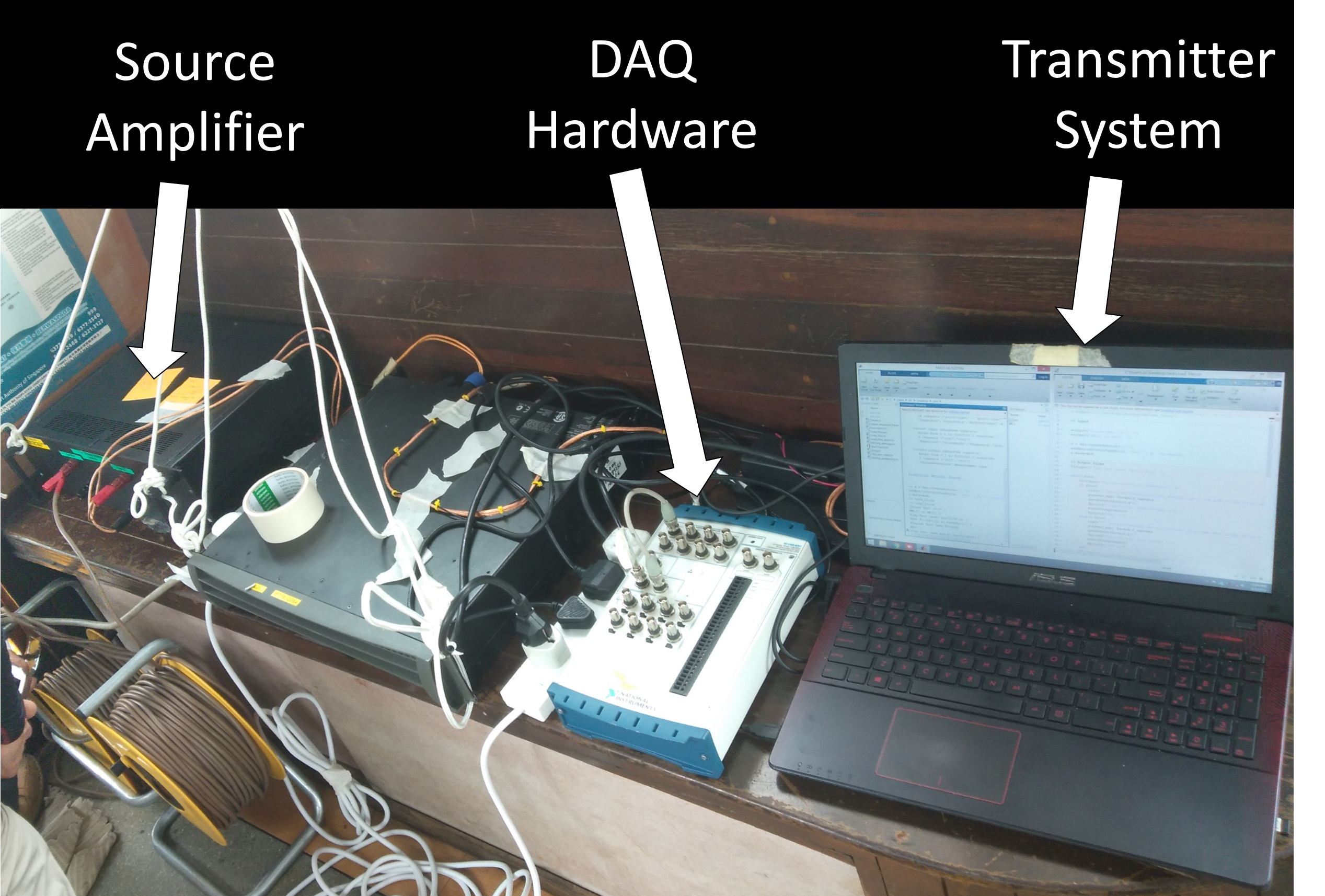}
        \caption{Transmitter Set-up}
        \label{Transmitter Set-up}
    \end{subfigure}\\
    \begin{subfigure}[b]{0.4\textwidth}
    \centering
        \includegraphics[width=\textwidth]{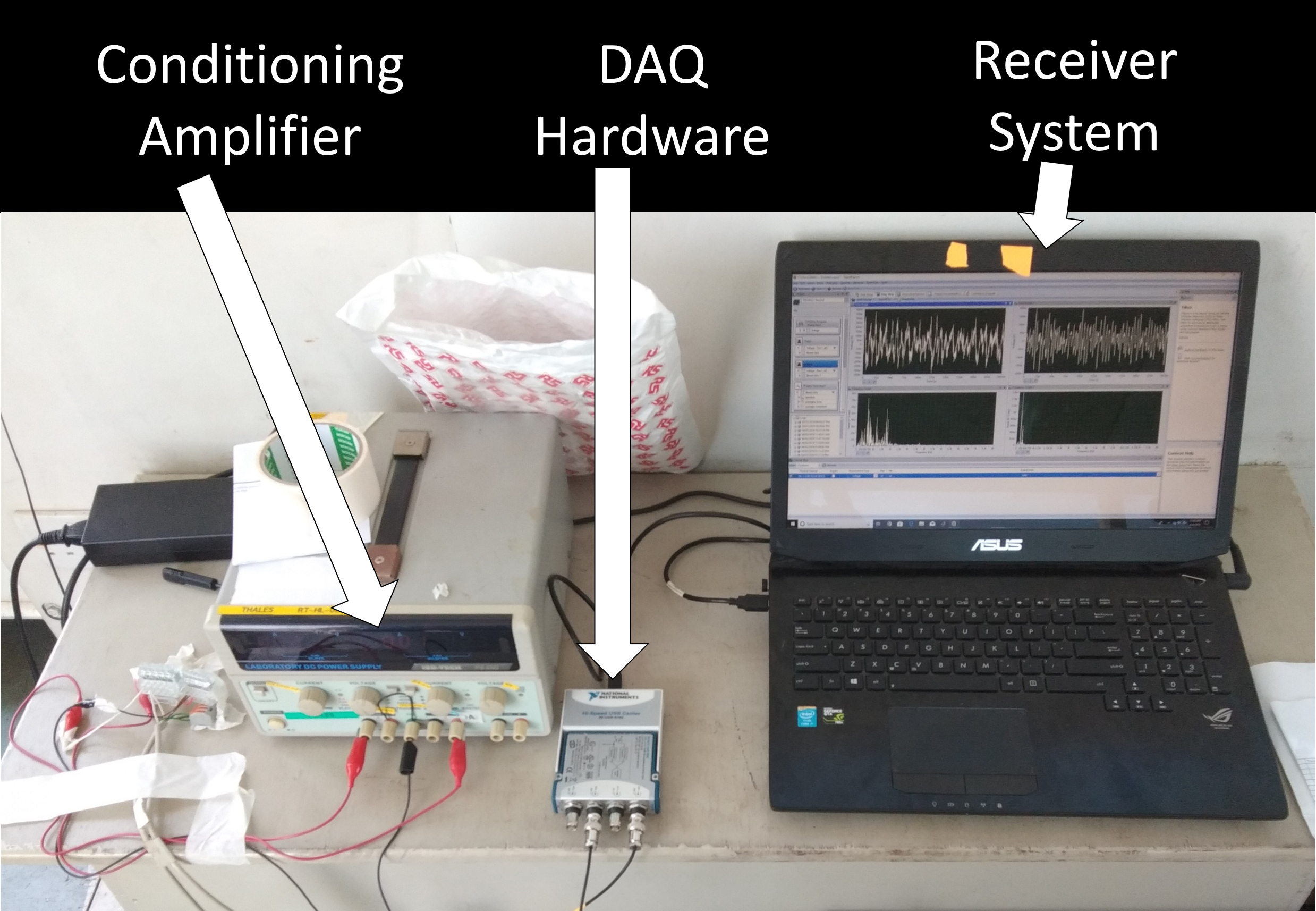}
        \caption{Receiver Set-up}
        \label{Receiver Set-up}
    \end{subfigure}\\
    \caption{Sea Trial On-site Set-up}
    \label{Sea Trial Experiment On-site Set-up}
\end{figure}

The communication system used in the underwater acoustic sea trial is depicted in Fig.\ref{Sea Trial End-to-End System Diagram}. Before transmission, the desired transmitted signal $x(t)$ is converted from digital values to analog sensor signals using a National Instruments-Data Acquisition (NI-DAQ) hardware unit. The signal is then amplified before being transmitted. At the receiver end, the signal is first received by the hydrophone and amplified by ISO-TECH IPS-3303. The corresponding NI-DAQ will translate the analog sensor signal to digital values for the proposed communication system. 

In March 2019, a sea trial was conducted in the waters near Selat Pauh, Singapore, where the bottom is muddy with the deepest depth of approximately 25m. The waters is considered to be relatively stationary with occasional disturbance from the large vessels traveling to the port. In this trial, the distance and depth of the transmitter and receiver was kept at about 300m and 9m respectively, with a variation of 50m and 1m due to the changing currents. The carrier frequency of the BPSK modulated signal was varied at 1kHz intervals for different trials. The trials were conducted at $f_c=1$kHz, $f_c=2$kHz, $f_c=3$kHz and $f_c=4$kHz. Due to the limitations of the hardware used in the trial, the sampling rate was set at 40 kHz. The specifications of the sea trial is recorded in Table.\ref{Specifications on the Sea Trial Experiment}. 

\begin{table}[h]
\centering
\caption{Specifications on the Sea Trial}
\label{Specifications on the Sea Trial Experiment}
\scalebox{1}{%
\begin{tabular}{l c}
\toprule
Trial Parameters & Value\\ \midrule
Distance between Transducer \& Hydrophone & 300m $\pm$ 50 m \\ 
Depth of Transducer &  9m $\pm$ 1 m \\ 
Amplifier Gain of Transducer &  25dB \\ 
Depth of Hydrophone & 9m $\pm$ 1 m \\ \bottomrule
\end{tabular}}
\end{table}

\subsection{Sea Trial Results}

In this subsection, the collected data from the sea trial described in Section \ref{Experimental Set-up} was used to validate the real application of the proposed receiver system. The results of which are shown in Table.\ref{Accuracy comparison of the method with real sea data}. 

\begin{table}[h]
\centering
\caption{Sea Trial Data Results and Accuracy Comparison between MLE with Doppler Synchronization and Proposed Receiver System}
\label{Accuracy comparison of the method with real sea data}
\scalebox{0.8}{%
\begin{tabular}{l c c c c c}
\toprule
\begin{tabular}[c]{@{}c@{}}Trial\\ No.\end{tabular} & $f_c$ & SNR & $\alpha$ & \begin{tabular}[c]{@{}c@{}}BER of MLE\\ + Doppler Sync.\end{tabular} & \begin{tabular}[c]{@{}c@{}}BER of Proposed\\ Receiver System\end{tabular}\\ \toprule
1 & 4kHz & -6.081 & 1.02 & 0.485 & 0.0148 \\ 
2 & 2kHz & -1.845 & 1.00 & 0.435 & 0.0330\\ 
3 & 3kHz & -4.818 & 1.12 & 0.490 & 0.0093\\ 
4 & 1kHz & -4.638 & 1.00 & 0.465 & 0.0102\\ \midrule
5 & 4kHz & -21.409 & 1.09 & 0.535 & 0.0790\\ 
6 & 4kHz & -22.019 & 0.90 & 0.495 & 0.0857\\ \midrule
7 & 4kHz & -28.468 & 0.87 & 0.492 & 0.0993\\ 
8 & 2kHz & -23.466 & 1.01 & 0.486 & 0.0899\\ 
9 & 1kHz & -28.781 & 1.14 & 0.502 & 0.1240\\ 
10 & 1kHz & -24.951 & 0.99 & 0.512 & 0.0917\\ \bottomrule
\end{tabular}}
\end{table}

The estimated average $\alpha$ was calculated using the up and down sweep of the HFM pilot signal and the SNR was estimated using MATLAB. For this evaluation, we collected data for 10 trials. Trial 1-4 were conducted on Day 1 and the results obtained from the sea trial were significantly better than seen in Fig.\ref{Simulation under sea trial conditions} with the most significant on Exp.3 with an coded BER of 0.0093 in comparison to BER of 0.045. This implies that during Day 1, the complexity of the channel was significantly lower than that simulated in the above trial. The data collected from Trials 5 and 6 on Day 2 had a much lower performance significance with an improvement of 0.08. On Day 3, while carrying out Exp. 7-10, we experienced heavy rain, which resulted in a more complex dataset. As such, the BER seen from the sea trials conducted on Day 3 shows similar performance to the simulated results. Overall, our proposed receiver system is able to keep a significant performance improvement from the $10^{-1}$ BER of the MLE with Doppler sync. to a $10^{-2}$ BER.

\section{Conclusion and Outlook}
\label{S5}
In this paper, we have proposed a novel receiver system that uses DBNs to redesign the de-noising and demodulation techniques for underwater acoustic communications. Our approach has also provided an interesting and important pathway for the application of machine learning techniques to underwater communications systems. 

Firstly, although the performance of the receiver system matches performance of traditional systems, without significant improvement, in the AWGN channels, it does show better performance in the more realistically simulated underwater channels influenced by Doppler and multi-path. A comparison with the traditional MLE and the promising de-noising auto encoder was completed in various underwater scenarios. These simulated experiments revealed extremely competitive BER performances with a performance improvement of 13.2dB at $10^{-3}$ BER. Therefore, demonstrating the powerful potential for machine learning to be used in more complex underwater acoustic channels. As a further investigation, we will increase the complexity by accommodating different mixtures of noise like rayleigh noise and exponential noise. 

As an additional step, we collected real life data, through a sea trial, to analyze the performance of the proposed receiver in a real scenario. The results of which were promising with a substantial improvement from a coded $10^{-1}$ BER using the traditional MLE method with Doppler synchronization to a coded $10^{-2}$ BER with the proposed receiver. This implies the real possibility of designing machine learning based underwater acoustic communication systems.

Finally, the strength of using DBNs to design our proposed receiver is denoted by its seemingly learned ability to comprehend and classify differing sets of received signals. Despite the varying parameters-- frequency, amplitude, phase and time frames between each random shift-- of the scenarios we have chosen to examine the receivers under, our proposed receiver has remained relatively invariant with the largest variation of 5.2dB at an uncoded $10^{-4}$ BER between the presence of 1-path and 3-paths. This phenomena suggests that the DBNs have successfully extracted meaningful features from the signals that could potentially be unchanged to the fluctuations of the underwater channel. 

\appendices

\bibliographystyle{IEEEtran}
\bibliography{IEEEabrv,biblio_rectifier}

\end{document}